\documentclass[12pt, twoside]{article}
\usepackage{a4wide,amssymb,cite}
\usepackage{epsfig}

\newcommand{\be}{\begin{equation}}
\newcommand{\ee}{\end{equation}}
\newcommand{\bea}{\begin{eqnarray}}
\newcommand{\eea}{\end{eqnarray}}
\def\eq#1{eq.~(\ref{#1})}

\begin{document}

\begin{titlepage}

\rightline{CERN-PH-TH/2010-211}

\begin{centering}
\vspace{1cm}
{\large {\bf Unitarizing Higgs Inflation}} \\

\vspace{1.5cm}

 {\bf Gian F. Giudice} and {\bf Hyun Min Lee} \\
\vspace{.2in}

{\it CERN, Theory Division, CH-1211 Geneva 23, Switzerland} \\

\vspace{.1in}

\end{centering}
\vspace{2cm}

\begin{abstract}
\noindent
We consider a simple extension of the Standard Model Higgs inflation with one new real scalar field which preserves unitarity up to the Planck scale. The new scalar field (called sigma) completes in the ultraviolet the theory of Higgs inflation by linearizing the Higgs kinetic term in the Einstein frame, just as the non-linear sigma model is unitarized into its linear version. The unitarity cutoff of the effective theory, obtained by integrating out the sigma field, varies with the background value of the Higgs field.
In our setup, both the Higgs field and the sigma field participate in the inflationary dynamics, following the flat direction of the potential. We obtain the same slow-roll parameters and spectral index as in the original Higgs inflation but we find that the Hubble rate during inflation depends not only on the Higgs self-coupling, but also on the unknown couplings of the sigma field.

\end{abstract}

\vspace{3cm}
\begin{flushleft}

\end{flushleft}

\end{titlepage}

\section{Introduction}

Inflation is believed to be the phenomenon that determined the necessary initial conditions for the cosmological evolution of our universe. Although there is mounting observational evidence in favor of inflation, the nature of the inflaton is still a mystery and a compelling link with an established particle theory is still missing. An interesting proposal that aims at filling this gap between cosmology and particle physics is the idea that the Standard Model Higgs boson could play the role of the inflaton~\cite{higgsinf} (see also ref.~\cite{others1}-\cite{others3}).  This appears to be possible if a Higgs bilinear term is coupled to the scalar curvature with an unusually large constant $\xi$ of the order of $10^4$. At first sight, this scenario suffers from a potential problem. Because of the large coupling constant $\xi$, the theory violates unitarity at the energy $M_P/\xi$. This energy is comparable to the inflationary Hubble rate and is parametrically smaller than the scale of the Higgs field during inflation, which is as low as $M_P/\sqrt{\xi}$. The violation of unitarity~\cite{unitarybound,mcdon} at the scale $M_P/\xi$ occurs in the theory expanded around
the vacuum in which the Higgs field takes a small value, of the order of the electroweak scale.
But, as emphasized in ref.~\cite{bmss} (see also ref.~\cite{ferrara}), this result does not necessarily spoil the self-consistency of the Higgs inflationary scenario. The energy cutoff, dictated by unitarity arguments, is field dependent. While being equal to $M_P/\xi$ at small field value, the energy cutoff grows as the Higgs background field is increased. Thus,
the region of the scalar potential relevant during the inflationary epoch could be within the domain of calculability.  

Nonetheless, there are reasons to be concerned with the unitarity issue of the theory. 
First of all, the cutoff associated with the would-be Goldstone bosons in
the Higgs doublet is lower than the one read from the potential of the Higgs boson. In unitary gauge, the problem becomes manifest in the gauge sector and it has been shown \cite{bmss} that the cutoff during inflation is given by $M_P/\sqrt{\xi}$, which is parametrically close to the energy scales involved during inflation.
Moreover, if we assume that the Higgs theory is eventually embedded into a more complete scheme that can be reliably extrapolated all the way up to the Planck mass, we will necessarily find new particles or new dynamics appearing at the scale $M_P/\xi$. It is quite reasonable to expect that the new degrees of freedom with mass of the order of $M_P/\xi$ will affect the Higgs potential at these scales and modify its form for values of the Higgs field relevant for inflation, of the order of $M_P/\sqrt{\xi}$. 
If this is the case, any assessment about the viability of Higgs inflation will require knowledge of the physics responsible for unitarization at the scale $M_P/\xi$. 

In this paper we will address the issue of unitarization of Higgs inflation. We will present a simple extension of the model, with only one new scalar field, that allows to raise the unitarity cutoff up to the Planck mass. The inflationary process, which can then be reliably computed, occurs in a fashion analogous to the original Higgs model. The procedure for unitarization that we follow is very similar to the one that is used to promote the non-linear sigma model into its linear version. The new scalar field that we introduced, called $\sigma$, plays the role of nearly linearizing the non-renormalizable Higgs interactions, as explained in sect.~2. 

The paper is organized as follows.
First, we sketch the procedure for unitarizing Higgs inflation and provide a simple model that implements this scheme with one scalar field.
Then we discuss the dynamics of the $\sigma$ field both in the vacuum and during the inflationary regime.
In sect.~3 we calculate the slow-roll parameters in our model and compare them to the original Higgs inflation.
To prove unitarity of our model in sect.~4 we consider the gauge-Higgs interactions and the Yukawa couplings and study their impact. Finally, some conclusions are drawn. There are two appendices dealing with the formalism for multi-field inflation and the details of the calculation of the slow-roll parameters in our model.

\section{The model}

In this section, we first explain the procedure for unitarizing Higgs inflation with a new real-scalar field.
Then we consider a simple model realizing this scheme with general renormalizable interactions and non-minimal couplings to gravity for the Higgs doublet and the new scalar field. Next we discuss the vacuum structure and the inflationary dynamics along the flat direction in our model.

\subsection{The procedure for unitarizing Higgs inflation}

The original model of Higgs inflation is based on the Jordan-frame Lagrangian~\cite{higgsinf}
\be
\frac{{\cal L}_J}{\sqrt{-g_J}}=\frac{1}{2} \Big(M^2_P+2\xi_0 {\cal H}^\dagger{\cal H}\Big)R-|D_\mu{\cal H}|^2-\lambda \Big({\cal H}^\dagger {\cal H}-\frac{v^2}{2}\Big)^2,
\ee
where $\xi_0$ is the non-minimal coupling of the Higgs doublet. The Einstein-frame Langrangian is obtained after the Weyl rescaling $g^J_{\mu\nu}=f g^E_{\mu\nu}$ with $f=(1+2\xi_0{\cal H}^\dagger {\cal H}/M^2_P)^{-1}$,
\bea
\frac{{\cal L}_{E}}{ \sqrt{-g_E}}&=&\frac{1}{2}M^2_P R-  \frac{|D_\mu {\cal H}|^2}{1+2\xi_0{\cal H}^\dagger {\cal H}/M^2_P} \nonumber \\
&&-\frac{3\xi^2_0}{M^2_P}\frac{\partial_\mu ({\cal H}^\dagger {\cal H})\partial^\mu ({\cal H}^\dagger {\cal H})}{(1+2\xi_0{\cal H}^\dagger {\cal H}/M^2_P)^2}
- \frac{ \lambda \Big({\cal H}^\dagger {\cal H}-\frac{v^2}{2}\Big)^2}{(1+2\xi_0{\cal H}^\dagger {\cal H}/M^2_P)^2} \, .\label{ehiggslimit}
\eea
This form of the Lagrangian clearly exhibits the unitarity problem in Higgs inflation, which originates from the first term in the second line of \eq{ehiggslimit}. The non-renormalizable dimension-6 operator involving four Higgs fields and two derivatives is suppressed by the mass scale $M_P/\xi_0$, which plays the role of the energy cutoff at small field value.

A procedure for unitarization is suggested by the analogy between the
Einstein-frame kinetic term for the Higgs and the non-linear sigma model, which is more transparent in the real representation with ${\cal H}^T=\frac{1}{\sqrt{2}}(\phi_1,\phi_2,\phi_3,\phi_4)$,
\be
{\cal L}_{\rm kin}=-\frac{1}{2(1+\xi_0{\vec\phi}^2/M^2_P)}\bigg(\delta_{ij}+\frac{6\xi^2_0\phi_i\phi_j/M^2_P}{1+\xi_0{\vec\phi}^2/M^2_P}\bigg)\partial_\mu\phi_i\partial^\mu\phi_j
\ee
with ${\vec\phi}^2\equiv\sum_i \phi^2_i$.
Just as a non-linear sigma model can be completed in the ultraviolet by the presence of a sigma field, 
we introduce a real-scalar sigma field satisfying the constraint $\sigma^2=\Lambda^2+{\vec\phi}^2$, with 
$\Lambda^2\equiv {M^2_P}/{\xi_0}$, and rewrite the Higgs kinetic term in the form 
\be
{\cal L}_{\rm kin}=-\frac{1}{2}\Big(\frac{\Lambda}{\sigma}\Big)^2\Big[(\partial_\mu\phi_i)^2
+6\xi_0(\partial_\mu\sigma)^2\Big]-\kappa (x)F(\sigma^2-\Lambda^2-{\vec\phi}^2) . \label{linearsig}
\ee
Here $\kappa (x)$ is the Lagrange multiplier and $F$ is an arbitrary function 
satisfying $F(0)=0$. The Higgs kinetic term does not yet correspond to a flat metric of the target space,
but rather it looks similar to the metric of Euclidean AdS$_5$ space with AdS radius $1/\Lambda$. However, as suggested by the constraint for the sigma field, it is possible to complete the theory into a linear sigma-model type, in which the sigma field vev is dynamically determined to be $(\Lambda^2+{\vec\phi}^2)^{1/2}$ by the full potential. This effectively corresponds to replacing the Lagrange-multiplier term by an appropriate scalar potential whose minimum lies at the field value $\sigma^2=\Lambda^2+{\vec\phi}^2$.
Then, after the field redefinition $\sigma = \Lambda \exp[\chi / (\sqrt{6} M_P)]$, we find that the canonically-normalized field $\chi$ has only Planck-suppressed non-renormalizable interactions. This allows to raise the unitarity cutoff up to $M_P$. 

In ref.~\cite{mcdonald} it was claimed that Higgs inflation could be unitarized by introducing additional non-renormalizable operators in the Jordan frame with their coefficients carefully chosen to cancel exactly the dangerous interactions causing the loss of unitarity. We believe that a dynamical solution is necessary to solve the problem. In the next section we will propose a simple model that implements our procedure for unitarization, completing Higgs inflation in the ultraviolet. 

\subsection{Higgs inflation with the sigma field}

Our model, which extends the original Higgs inflation by adding to the SM Higgs doublet ${\cal H}$ a real scalar $\bar\sigma$, is based on the Jordan-frame Lagrangian 
\bea
\frac{{\cal L}_J}{\sqrt{-g_J}}&=&\frac{1}{2} \Big({\bar M}^2+\xi{\bar\sigma}^2+2\zeta {\cal H}^\dagger{\cal H}\Big)R-\frac{1}{2}(\partial_\mu{\bar\sigma})^2-|D_\mu{\cal H}|^2 \nonumber \\
&&-\frac{1}{4}\kappa \Big({\bar\sigma}^2- {\bar\Lambda}^2-2\alpha{\cal H}^\dagger {\cal H}\Big)^2-\lambda \Big({\cal H}^\dagger {\cal H}-\frac{v^2}{2}\Big)^2. \label{jordanaction0}
\eea
Here $\bar M$, $\bar \Lambda$, and $v$ are parameters with dimension of mass. We assume that the electroweak scale $v$ is much smaller than the other masses involved in the Lagrangian ($v\ll \bar M,\bar \Lambda$). This assumption, technically unnatural, is just an expression of the hierarchy problem, which cannot be addressed in the SM using conventional symmetry arguments. Since we are working in the context of the SM, we must accept this assumption without a known justification. In \eq{jordanaction0}, the parameters
$\xi,\zeta$ are the non-minimal couplings of the sigma field and the Higgs doublet to the scalar curvature. As described later, inflation requires a large coupling $\xi$, of the order of $10^4$. On the other hand, we will take $\zeta$ of order unity, in order to avoid the reappearance of the unitarity problem in the Higgs sector. It is technically unnatural to set $\zeta =0$, because $\zeta$ can be generated by loop effects, but it is possible to keep it significantly smaller than $\xi$. We assume that this is the case. Finally 
 $\kappa,\alpha, \lambda$ are dimensionless coupling constants. The Lagrangian (\ref{jordanaction0}) contains the most general renormalizable terms compatible with the $Z_2$ symmetry under which $\bar\sigma$  transforms as ${\bar\sigma}\rightarrow -{\bar\sigma}$. 
 
It is useful to choose the unitary gauge for the Higgs doublet, ${\cal H}^T=\frac{1}{\sqrt{2}}(0,\phi)$, and introduce
the field variable $\sigma$ with the definition $\sigma^2={\bar\sigma}^2+M^2$ with $M^2\equiv {\bar M}^2/\xi$. With this transformation, the above Lagrangian can be rewritten in a form in which the role of the Planck mass is expressed only in terms of fields,
\bea
\frac{{\cal L}_J}{\sqrt{-g_J}}=\frac{1}{2}(\xi\sigma^2+\zeta  \phi^2)R-\frac{\sigma^2}{2(\sigma^2-{M}^2)}(\partial_\mu\sigma)^2-\frac{1}{2}(\partial_\mu \phi)^2-V_J
\label{jordanaction}
\eea
where
\be
V_J=\frac{1}{4}\kappa \left(\sigma^2- \Lambda^2-\alpha \phi^2\right)^2
+\frac{\lambda}{4} \left(\phi^2-v^2\right)^2 \label{jordanpot}
\ee
and $\Lambda^2\equiv M^2+{\bar\Lambda}^2$.
Thus, the Planck mass is traded off for the non-canonical kinetic term for the new sigma field in Jordan frame.
Since the minimization of the potential sets $\langle \sigma \rangle = \Lambda$ (up to negligible corrections of order $v^2$), the field $\sigma$ determines the effective Planck mass. 
So, we need to choose 
\be
\Lambda=\frac{M_P}{\sqrt{\xi}}.
\ee

We can now rewrite the Lagrangian in the Einstein frame by performing a Weyl rescaling of the metric, $g^J_{\mu\nu}=f g^E_{\mu\nu}$ with $f=M_P^2/(\xi \sigma^2+\zeta \phi^2)$,
\bea
\frac{{\cal L}_E}{\sqrt{-g_E}}&=&\frac{1}{2}M^2_P R 
-\frac{M_P^2}{2(\xi \sigma^2 +\zeta \phi^2)}\left[ \left( \frac{\sigma^2}{\sigma^2-M^2}+\frac{6\xi^2\sigma^2}{\xi \sigma^2 +\zeta \phi^2}\right) (\partial_\mu \sigma )^2 
\right. \nonumber \\  && \left.
+ \left( 1+\frac{6\zeta^2\phi^2}{\xi \sigma^2 +\zeta \phi^2}\right) (\partial_\mu \phi )^2 
+\frac{3\xi \zeta}{2(\xi \sigma^2 +\zeta \phi^2)}\partial_\mu \sigma^2 \partial_\mu \phi^2 \right] -V_E \label{einlag55}
\eea
\be
V_E=\frac{M_P^4}{4(\xi \sigma^2 +\zeta \phi^2)^2}\left[ \kappa \left(\sigma^2- \Lambda^2-\alpha\phi^2\right)^2
+\lambda (\phi^2-v^2)^2\right]. \label{einpot2}
\ee
Note that the field $\sigma$ is such that $\sigma^2 > M^2$, because of its definition in terms of $\bar \sigma$, and thus the sign of the kinetic term for $\sigma$ is well defined and no ghost-like instabilities exist. 

In the limit $\zeta =0$ and $M=0$ the Lagrangian in \eq{einlag55} exhibits a form similar to
the one in eq.~(\ref{linearsig}), suggested by the sigma-model discussion, apart from the coefficient of the sigma-field kinetic term which is $(1+6\xi)$ instead of $6\xi$.
In our case, the scalar potential $V_E$ contains the term playing the role of the Langrange multiplier in setting the constraint $\sigma^2 = \Lambda^2+\alpha\phi^2$. The limit $M=0$ corresponds to the case of induced gravity~\cite{induced}. In the limit $\alpha =0$ the theory has strong similarities with the model proposed in ref.~\cite{russ}.

\subsection{Dynamics with the sigma field}

Let us now study the structure of the theory in the Einstein frame. The vacuum of the model lies at
\be
\langle \phi \rangle^2 = v^2, ~~~~\langle \sigma \rangle^2 = \Lambda^2 +\alpha v^2.
\ee
For $v\ll \Lambda$ and for large $\xi$, the kinetic mixing between $\sigma$ and $\phi$ in \eq{einlag55} becomes negligible and the fields $\chi =\sqrt{6}M_P\ln (\sigma /\Lambda)$ and $\phi$ are approximately canonically normalized. The mass of $\chi$ can then be read off from the potential in \eq{einpot2}, with the result
\be
m_\chi \simeq \sqrt{\frac{\kappa}{3}} \frac{M_P} {\xi}.
\ee
As expected, the mass of the new degree of freedom described by the field $\sigma$ turns out to be of the order of $M_P/\xi$, the energy scale at which the original Higgs model violates unitarity. Below the scale $M_P/\xi$ we can integrate out the field $\sigma$ and obtain an effective theory, which corresponds to the original Higgs inflation model. Up to some higher-dimensional terms suppressed by $M_P/\sqrt{\xi}$, the effective theory is described by  the Lagrangian (\ref{ehiggslimit}) with $\xi_0 =\alpha \xi +\zeta$.  Above the scale $M_P/\xi$, the sigma field cures the unitarity breakdown of the original Higgs inflation, as is easily understood by replacing $\sigma$ in the Lagrangian of \eq{einlag55} with its expression in terms of the $\chi$ field, $\sigma = \Lambda \exp (\chi /\sqrt{6} M_P)$. All the non-renormalizable interactions are suppressed by the Planck mass.

Let us now consider the theory for large values of the Higgs background. 
For $|\sigma|, |\phi |\gg \Lambda$, the Einstein-frame potential (\ref{einpot2}) becomes approximately a function of only the ratio between $\phi$ and $\sigma$. It is then convenient to rewrite the Lagrangian (\ref{einlag55}) in terms of the field ${\tilde \phi} = \Lambda \phi /\sigma$ and obtain
\bea
\frac{{\cal L}_E}{\sqrt{-g_E}}&=&\frac{1}{2}M^2_P R 
-\frac{1}{2(1+\zeta{\tilde\phi}^2/M^2_P)}\left\{ \left[ \frac{\Lambda^2\sigma^2}{\sigma^2-M^2}+(1+6\zeta){\tilde\phi}^2+6M_P^2\right] \left( \frac{\partial_\mu\sigma}{\sigma}\right)^2 \right.\nonumber \\ && \left.
+\frac{1+\zeta(1+6\zeta){\tilde\phi}^2/M^2_P}{1+\zeta{\tilde\phi}^2/M^2_P}(\partial_\mu {\tilde\phi})^2 +(1+6\zeta)
\frac{\partial_\mu\sigma}{\sigma}\partial^\mu{\tilde\phi}^2\right\} -V_E. \label{einsteinaction1}
\eea
At the leading order, the potential for $\tilde \phi$ is
\be
V_E\simeq \frac{\Lambda^4}{4(1+\zeta{\tilde\phi}^2/M^2_P)^2}\bigg[(\lambda+\kappa\alpha^2)\Big(\frac{\tilde\phi}{\Lambda}\Big)^4-2 \kappa\alpha\Big(\frac{\tilde\phi}{\Lambda}\Big)^2+\kappa \bigg],
\label{pott}
\ee
whose minimum is
at ${\tilde\phi}\simeq\sqrt{\frac{\kappa\alpha}{\lambda+\kappa\alpha^2}}\Lambda$ for $\frac{\zeta}{\xi}\ll 1$. 
 Once $\tilde\phi$ is frozen at its minimum value, the potential presents a flat direction along the field component orthogonal to $\tilde \phi$. From \eq{einsteinaction1} we find that, during inflation, the fields  $\tilde \phi$ and
$\chi =\sqrt{6}M_P\ln (\sigma /\Lambda)$ are approximately canonically normalized and their kinetic mixing is negligible. The scalar potential for $\chi$ is obtained by keeping higher orders in $\Lambda^2$ in \eq{einpot2} and by freezing $\tilde\phi$ into its minimum. Ignoring terms proportional to $\zeta/ \xi$, we then find
\be
V_E\simeq V_{\rm inf} \left( 1-2 e^{-\frac{2\chi}{\sqrt{6}M_P}}\right) ,~~~~
V_{\rm inf}\equiv  \frac{\Lambda^4}{4}\Big(\frac{\lambda\kappa}{\lambda+\kappa\alpha^2}\Big) .
\label{eq15}
\ee 
The potential $V_E$ along the $\chi$ direction is exponentially flat, in perfect analogy with the original Higgs inflation. 

Note that the mass of the heavy mode $\tilde \phi$ during inflation is equal to $m_{\tilde \phi} \simeq \sqrt{2\kappa \alpha}~ \Lambda$, which is of the order of $M_P/\sqrt{\xi}$. Therefore, the mass of the heavy mode, which is about $M_P/\xi$ in the vacuum, is raised to $M_P/\sqrt{\xi}$ when the fields obtain large background values. In the original Higgs inflation, although new dynamics has to appear at the mass scale $M_P/\xi$ to cure the unitarity problem, during inflation the energy cutoff is higher, and is equal to about $M_P/\sqrt{\xi}$. Thus, our model gives an explicit realization of the mechanism advocated in ref.~\cite{bmss}. In our model, the inclusion of the field $\sigma$ allows for full control of the theory up to the Planck scale.

\section{Slow-roll inflation}

As discussed in the previous section, our Higgs inflation model with the sigma field is reduced to a single field inflation along the flat direction. In this section, 
we show that the flat direction indeed drives inflation by explicitly computing the slow-roll parameters. In Appendices A and B, we perform the calculation using the general formalism for multi-field inflation and we show that, in our case, the single-field approximation is adequate. We are interested in the case in which the non-minimal coupling of the Higgs doublet is much smaller than the one of the sigma field. 

The $\varepsilon$ slow-roll parameter, see eq.~(\ref{epsilon1}),
for $|\sigma|\gg \Lambda$ is given by
\be
\varepsilon\simeq \frac{4}{3}\Big(\frac{\Lambda}{\sigma}\Big)^4. \label{epsilon}
\ee
The differential number of e-foldings, see eq.~(\ref{nefold}), is $dN\simeq -\frac{\partial_\sigma V_E}{2\varepsilon V_E}$ for $\frac{\partial V_E}{\partial{\tilde \phi}}=0$ along the flat direction, so the total number of e-foldings is given by
\be
N\simeq-\int^{\sigma_f}_{\sigma_i}\frac{1}{2\varepsilon V_E}\frac{\partial V_E}{\partial\sigma}d\sigma 
\simeq  \frac{3}{4}\left( \frac{\sigma_i^2}{\Lambda^2}-\frac{\sigma_f^2}{\Lambda^2} \right). \label{efolds0}
\ee
Here we take $\sigma_f^2=(2/\sqrt{3})\Lambda^2$, corresponding to the field value at which $\varepsilon =1$ and the dynamics exit the slow-roll regime.

The $\eta$ parameter is given by the minimum between the two expressions $\eta_1$ and $\eta_2$, corresponding to the two independent field directions.
For $|\sigma|\gg \Lambda$, from eqs.~(\ref{eta01}) and (\ref{eta02}), we obtain
\bea
\eta_1&\simeq & -\frac{4}{3}\Big(\frac{\Lambda}{\sigma}\Big)^2,  \label{eta1} \\
\eta_2&\simeq &  8\xi\alpha\Big(1+\frac{\kappa}{\lambda}\alpha^2\Big). \label{eta2}
\eea
Consequently, we find that the mass of the heavy field orthogonal to the flat direction is 
$m^2_2\simeq \eta_2 V_{\rm inf}/{M^2_P}\simeq 2\kappa\alpha\Lambda^2$, in agreement with the result in the previous section.
On the other hand,  the $\eta$ parameter for inflaton is given by $\eta= \eta_1$, so the inflaton mass is $m^2_1\simeq \eta_1 H^2$.
Combining eqs.~(\ref{epsilon}), (\ref{efolds0}), and (\ref{eta1}), we obtain the slow-roll parameters in terms of the number of e-foldings as 
\be
\varepsilon\simeq \frac{3}{(2N+\sqrt{3})^2}, ~~~~ \eta\simeq -\frac{2}{2N+\sqrt{3}}.
\ee

Since $d\ln k=dN$ for an approximately constant Hubble parameter during inflation, using eq.~(\ref{powerspec}) and $\partial_\sigma\varepsilon\simeq (\partial_\sigma\ln V_E) (-2\varepsilon+\eta)$ we obtain the spectral index 
\be
n_s\simeq 1+2\eta - 6\varepsilon \simeq 1-\frac{2(4N+9+2\sqrt{3})}{(2N+\sqrt{3})^2}.
\ee
The combined WMAP 7-year data with Baryon Acoustic Oscillations and Type Ia supernovae \cite{wmap7} show that the spectral index is $n_s=0.963\pm0.012$ ($68\%$ CL).
For $N\simeq 60$, we obtain $\varepsilon\simeq 2.0\times 10^{-4}$ and $\eta\simeq -1.6\times 10^{-2}$, leading to the spectral index $n_s\simeq 0.966$ and ratio of the tensor to scalar perturbations, $r= 12.4\, \varepsilon\simeq 2.4\times 10^{-3}$, both compatible with observations. All these results for the slow-roll parameters, number of e-foldings, and spectral index are identical to those of the original Higgs inflation.

The COBE normalization of the power spectrum constrains the inflation parameters 
\be
\frac{V^{1/4}}{\varepsilon^{1/4}}\simeq 6.7\times 10^{16}~ {\rm GeV}.\label{cobe}
\ee
For the vacuum energy during inflation $V_{\rm inf}$ in \eq{eq15}, the COBE normalization (\ref{cobe}) leads to 
\be
\xi\sqrt{\frac{\lambda+\kappa\alpha^2}{\kappa\lambda}}\simeq 5\times 10^4.
\ee
This determines the scale $\Lambda$ to be about $10^{16}$~GeV. This result suggests the interesting possibility that the vev of the singlet field $\sigma$ could be responsible for the scale of the right-handed neutrino masses.

The constraint on the non-minimal coupling $\xi$ of the sigma field depends on all the dimensionless parameters of our model.  On the other hand,
in the original Higgs inflation, the COBE normalization gives ${\xi_0}/{\sqrt{\lambda}}\simeq 5\times 10^4$. Here $\xi_0$ is the non-minimal coupling  of the Higgs doublet, which is related to the parameters of our model by  $\xi_0=\alpha\xi$, considering the effective theory with $\sigma$ integrated out. Therefore the constraint on $\xi$ coincides with the one of the original Higgs inflation only when $\kappa \alpha^2 \gg \lambda$. In general, however, it depends on the values of the various unknown coupling constants and cannot be simply related to the observable Higgs quartic coupling.

There is another important difference between the Higgs inflation and our model. In our case, at the end of inflation the field configuration will be at 
$\phi /\sigma \simeq \sqrt{\frac{\kappa \alpha } {\lambda+\kappa \alpha^2}}$ as discussed below eq.~(\ref{pott}). Therefore the inflaton is a combination of the Higgs and sigma fields with a mixing angle determined by the values of the various coupling constants. This mixing angle suppresses the decay of the inflaton, because only the Higgs is directly coupled to the SM particles. Thus, the reheating temperature in our case is smaller than the one in Higgs inflation, which is estimated to be about $10^{13}$~GeV~\cite{btrh}.

Moreover, it has been observed in ref.~\cite{run} that, in Higgs inflation, the loop corrections to the Higgs self-coupling are important for determining the spectral index with a precision measurable by PLANCK. In our case, see eq.~(\ref{einlag55}), the Higgs kinetic term is close to a canonical form, independently of  the background field values and 
so, it would be sufficient to consider the SM running of the Higgs self-coupling. However, the dependence on the running effect coming from the sigma field interactions prevents us from making a simple testable prediction.

\section{Gauge and fermion interactions}

The analysis of the scalar potential performed in sect.~2.3 has shown that no unitarity violation occurs below the Planck scale, independently of the background scalar field values. However, the choice of the unitary gauge hides part of the problem. The interactions of the would-be Goldstone bosons could introduce unitarity violations at a lower cutoff scale. This is actually happening in the case of the original Higgs inflation~\cite{unitarybound,bmss}.
In the unitary gauge, the extra degrees of freedom contained in the Higgs doublet are gauged away by a local gauge transformation and their information  is  encoded in the gauge-Higgs interactions. It is therefore important to analyze also the gauge and Yukawa couplings of the Higgs in order to check the absence of any unitarity violation.
In this section, we consider the power counting both in the true vacuum and in the inflationary background for gauge-Higgs interactions as well as Yukawa couplings and compare it to the original Higgs inflation. We again neglect contributions coming from the non-minimal coupling of the Higgs doublet with respect to the $\sigma$ coupling.

\subsection{Field fluctuations around the vacuum}

\begin{itemize}
\item {\bf Gauge interactions}

The gauge kinetic terms are conformally invariant under the Weyl rescaling of the metric.
So, the Higgs interactions with two gauge bosons in unitary gauge are given by 
\be
\frac{{\cal L}_{\rm gauge}}{\sqrt{-g_E}}= -\frac{1}{2}f g^2 \phi^2 A_\mu A^\mu \simeq -\frac{1}{2}\Big(\frac{\Lambda}{\sigma}\Big)^2
g^2\phi^2 A_\mu A^\mu. \label{gaugeint}
\ee
Expanding around the vacuum $\sigma=\Lambda+{\chi}/{\sqrt{6\xi}}$ and $\phi=v+h$,
the above gauge interaction becomes
\be
\frac{{\cal L}_{\rm gauge}}{\sqrt{-g_E}}\simeq-\frac{1}{2}\Big(1-\frac{2\chi}{\sqrt{6}M_P}\Big)
g^2v^2 \Big(1+\frac{h}{v}\Big)^2 A_\mu A^\mu.
\ee
Therefore, the Higgs-gauge interactions are identical to the SM and there is no unitarity violation, while the coupling of the sigma field to the gauge sector is suppressed by the Planck scale.  
This result is a direct consequence of the fact that the physical Higgs is not rescaled, contrary to the original Higgs inflation. In the original model, because of the correction to the Higgs kinetic term in the Einstein frame, the gauge-Higgs interactions are modified as compared to the SM: $-\frac{1}{2}g^2v^2 (1+2a \frac{h}{v} +b \frac{h^2}{v^2})A_\mu A^\mu$ with $a=1-\frac{3v^2}{\Lambda^2_{HI}}$ and $b=1-\frac{12v^2}{\Lambda^2_{HI}}$ where $\Lambda_{HI}=\frac{M_P}{\xi_0}$. So, the unitarity cutoff of the original Higgs inflation must be identified with $\Lambda_{HI}$.

\item {\bf Fermion interactions}

Let us consider the fermion kinetic terms in the Einstein frame
\be
\frac{{\cal L}_{\rm fermion}}{\sqrt{-g_E}}=f^{3/2} {\bar\psi}i\gamma^\mu\partial_\mu\psi\simeq\Big(\frac{\Lambda^2}{\sigma^2}\Big)^{3/2} {\bar\psi}i\gamma^\mu\partial_\mu\psi \, .
\ee
We can make the kinetic terms canonical by rescaling the fermions: $\psi'=(\frac{\Lambda^2}{\langle\sigma^2\rangle})^{3/4}\psi$.
The Yukawa couplings become
\bea
\frac{{\cal L}_{\rm Yukawa}}{\sqrt{-g_E}}&=& f^2 \lambda_\psi \phi\,{\bar\psi}_R\psi_L +{\rm h.c.} 
\nonumber \\
&\simeq&\left(\frac{\Lambda}{\sigma}\right)^2\left(\frac{\Lambda^2}{\langle\sigma^2\rangle}\right)^{-3/2}\lambda_\psi \phi \,{\bar\psi}'_R\psi'_L+{\rm h.c.} \label{yukawa}
\eea
Then, applying the same expansions of the scalar fields around the vacuum as for the gauge interactions,
the Yukawa couplings become
\be
\frac{{\cal L}_{\rm Yukawa}}{\sqrt{-g_E}}\simeq\Big(1-\frac{4\chi}{\sqrt{6}M_P}\Big)\lambda_\psi (v+h)
{\bar\psi}'_R\psi'_L+{\rm h.c.}
\ee
Again, the Yukawa couplings to the physical Higgs are the same as in the SM.
On the other hand, in the original Higgs inflation, there was a dimension-6 operator, $\lambda_\psi \frac{h^3}{\Lambda^2_{HI}}{\bar\psi}'_R\psi'_L$, which is suppressed by $\Lambda_{HI}$.

\end{itemize}

\subsection{Field fluctuations during inflation}

During inflation, the fields reside at large values, $|\sigma|\gg \Lambda$ and $\phi^2\simeq \frac{\kappa\alpha}{\lambda+\kappa\alpha^2}\sigma^2$. We can expand the scalar fields around the inflationary background as follows,
\be
\sigma\simeq \sigma_0\Big(1+\frac{1}{\sqrt{6\xi}\Lambda}\chi\Big),  \quad \phi\simeq\phi_0+\frac{\sigma_0}{\Lambda}h
\ee
where $\sigma_0,\phi_0$ are the background field values during inflation and $\chi,h$ are perturbations having canonical kinetic terms.

\begin{itemize}
\item {\bf Gauge interactions}

From eq.~(\ref{gaugeint}), the gauge interactions become
\bea
\frac{{\cal L}_{\rm gauge}}{\sqrt{-g_E}}
\simeq-\frac{1}{2}g^2V^2 \Big(1-\frac{2\chi}{\sqrt{6}M_P}\Big) \Big(1+\frac{h}{V}\Big)^2 A_\mu A^\mu
\eea
where $V\equiv {\Lambda\phi_0}/{\sigma_0} $. Thus, the Higgs-gauge interactions are of the standard form with the Higgs vev being replaced by $V$, so there is no unitarity violation in the gauge sector. Because of the large scalar values during inflation, the gauge boson mass is increased and saturates at 
$m_A\simeq gV\simeq g\Lambda\sqrt{\frac{\kappa\alpha}{\lambda+\kappa\alpha^2}} $.
However, even during inflation,  there is no unitarity violation below the Planck scale as in the vacuum case. In the original Higgs inflation, the gauge interactions are given by $-\frac{1}{2}g^2V^2_{HI}(1+\frac{h}{\sqrt{6}M_P})^2A_\mu A^\mu$ with $V_{HI}=\frac{M_P}{\sqrt{\xi_0}}$. Thus, due to the suppressed Higgs couplings, unitarity is broken at $V_{HI}$.

\item {\bf Fermion interactions}

From eq.~(\ref{yukawa}), the Yukawa interactions become
\bea
\frac{{\cal L}_{\rm Yukawa}}{\sqrt{-g_E}}
\simeq\Big(1-\frac{4\chi}{\sqrt{6}M_P}\Big)\lambda_\psi (V+h)
{\bar\psi}'_R\psi'_L+{\rm h.c.}
\eea
Thus, we find that the fermions have large masses but there is no unitarity violation below the Planck scale. 
The Yukawa couplings during inflation are not suppressed as compared to the SM ones,
unlike the original Higgs inflation in which the Yukawa interactions are given by $\lambda_\psi V_{HI} (1+\frac{h}{\sqrt{6}M_P}){\bar\psi}'_R\psi'_L$.

\end{itemize}

\section{Conclusions}

The idea that the Higgs boson could play the role of the inflaton is very intriguing. A scalar theory with quartic interaction in the potential and large non-minimal coupling $\xi$ to the curvature can support inflation. However, the inflationary dynamics occurs at such large values of the scalar field that the identification of the inflaton with the Higgs boson remains suspicious, in view of the existence of the intermediate scale $M_P/\xi$ at which the theory around its true vacuum violates unitarity. If we insist that the theory can be extended up to the Planck mass, it is quite plausible that the necessary new physics occurring at the scale $M_P/\xi$ will modify the Higgs potential in the regime relevant for inflation. Any conclusion about the viability of Higgs inflation will then require knowledge of the new dynamics that unitarizes the theory.

We have considered a simple model, with one additional scalar field $\sigma$, which cures the unitarity violation at the intermediate scale and allows for an extrapolation of the theory up to $M_P$. The procedure we followed to construct the model is reminiscent of the unitarization of the non-linear sigma model into its linear version. 

In our model, the $\sigma$ field has a mass of order  $M_P/\xi$ and the effective theory below this scale essentially corresponds to the original model of Higgs inflation~\cite{higgsinf}, namely the SM with a large non-minimal coupling between the Higgs and the curvature. The analysis of our model in the regime above $M_P/\xi$ shows that the theory can support inflation in a way completely analogous to the case of the original Higgs inflation. The predictions for the slow-roll parameters and the spectral index are identical in both theories. It is interesting that, in our model, the mass of the heavy mode increases with the field background. Being equal to $M_P/\xi$ around the true vacuum, the mass of the new state is about $M_P/\sqrt{\xi}$ during inflation, giving an explicit realization of the mechanism advocated in ref.~\cite{bmss}, for which the scale of unitarity violation is raised at large background field. In spite of the similarities with the original Higgs inflation, the $\sigma$ field plays a crucial role. Besides unitarizing the theory, $\sigma$ directly participates in the inflationary dynamics. Its role is also reflected in the fact that the relation between $\xi$ and the inflationary scale does not only depend on the measurable Higgs quartic coupling, but also on unknown coupling constants determining the $\sigma$ interactions. The reheating temperature in our model can be smaller than in the original Higgs inflation.

\section*{Acknowledgments}
We would like to thank F.~Bezrukov, C.~Burgess, A.~Linde, A. Riotto, M.~Shaposhnikov, S.~Sibiryakov and M.~Trott for comments and discussions.
H.M.L. is supported by the Korean-CERN fellowship.

\def\theequation{A.\arabic{equation}}
\setcounter{equation}{0}
\vskip0.8cm
\noindent
{\Large \bf Appendix A:  Multiple-field inflation}
\vskip0.4cm
\noindent

We review here the general formulas for slow-roll parameters and spectral index in inflation models with multiple scalars \cite{multifield}.

The Einstein-frame action with multi-scalars is 
\be
S_E=\frac{1}{2}\int d^4x \sqrt{-g_E} \left[M^2_P R -G_{IJ}\partial_\mu\varphi^I\partial^\mu\varphi^J -2V(\varphi)\right] .
\ee
Taking the metric $ds^2=-dt^2+a^2(t)\delta_{ij}dx^i dx^j$, and time-dependent scalars $\varphi^I$, the Einstein equation and the equation of motion of the scalars are
\bea
\Big(\frac{\dot a}{a}\Big)^2&=&\frac{1}{6M^2_P}\Big(G_{IJ}{\dot\varphi}^I{\dot\varphi}^J+2V\Big), \\
\frac{\ddot a}{a}&=&-\frac{2}{3M^2_P}(G_{IJ}{\dot\varphi}^I{\dot\varphi}^J-V), 
\eea
\bea
{\ddot\varphi}^I+3H{\dot\varphi}^I+\Gamma^I_{JK}{\dot\varphi}^J{\dot\varphi}^K+G^{IJ}V_{,J}=0.
\eea

The $\varepsilon$ slow-roll parameter for multi-field inflation is defined as
\be
\varepsilon=-\frac{\dot H}{H^2} = \frac{1}{2M^2_P H^2}\,G_{IJ}{\dot\varphi}^I{\dot\varphi}^J.
\ee
For ${\ddot\varphi}^I+\Gamma^I_{JK}{\dot\varphi}^J{\dot\varphi}^K\ll G^{IJ}V_{,J}$, we can rewrite
the above slow-roll parameter as
\be
\varepsilon\simeq \frac{M^2_P}{2V^2}\, G^{IJ}V_{,I} V_{,J}. \label{genepsilon}
\ee

The counterpart of the $\eta$ slow-roll parameter in multi-field inflation is defined as $\eta={\rm min}_a \eta_a$
where $\eta_a$ are eigenvalues of the matrix $N_I\,^J$, 
\be
N_I\,^J = M^2_P\,\frac{G^{JK} V_{;KI}}{V}  \label{generaleta}
\ee
where $V_{;IJ}\equiv \partial_I\partial_J V-\Gamma^K_{IJ}\partial_K V$.

The number of e-foldings is defined as
\be
dN=H dt = -\frac{1}{\varepsilon H} dH=-\frac{1}{\varepsilon H}\frac{\partial H}{\partial\varphi^I} 
d\varphi^I.
\ee
From the Einstein equations, we obtain
\be
{\dot H}=-\frac{1}{2M^2_P} G_{IJ}{\dot\varphi}^I{\dot\varphi}^J.
\ee
Thus, since ${\dot H}=\frac{\partial H}{\partial\varphi^I}{\dot\varphi}^I$, we get
\be
\frac{\partial H}{\partial\varphi^I}=-\frac{1}{2M^2_P} G_{IJ}{\dot\varphi}^J.
\ee
Then, for ${\ddot\varphi}^I+\Gamma^I_{JK}{\dot\varphi}^J{\dot\varphi}^K\ll G^{IJ}V_{,J}$, 
we get $\frac{1}{H}\frac{\partial H}{\partial\varphi^I}\simeq \frac{V_{,I}}{6H^2M^2_P}\simeq \frac{V_{,I}}{2V}$.
Therefore, we obtain the following approximate expression for $N$ 
\be
N\simeq-\int^{\varphi^I_f}_{\varphi^I_i} \frac{V_{,I}}{2\varepsilon V} d\varphi^I. \label{nefold}
\ee

The power spectrum for the multi-field inflation is given by
\be
P(k)=\frac{V}{75\pi^2M^2_P} G^{IJ}\frac{\partial N}{\partial\varphi^I}\frac{\partial N}{\partial\varphi^J}.
\ee
Using the approximate formula,
\be
\frac{\partial N}{\partial\varphi^I}= -\frac{1}{\varepsilon H} \frac{\partial H}{\partial\varphi^I}
\simeq -\frac{V_{,I}}{6M^2_P H^2\varepsilon},
\ee
and eq.~(\ref{genepsilon}),
we get the power spectrum as for the single-field inflation,
\be
P(k)\simeq \frac{V}{150\pi^2M^4_P\varepsilon}. \label{powerspec}
\ee
Finally, the spectral index is 
\be
n_s=1+\frac{\partial \ln P(k)}{\partial \ln k}.
\ee

\def\theequation{B.\arabic{equation}}
\setcounter{equation}{0}
\vskip0.8cm
\noindent
{\Large \bf Appendix B:  Calculation of slow-roll parameters}
\vskip0.4cm
\noindent

We apply the general formulas for multi-field inflation to calculate the inflationary observables in our model.  We consider the case in which the non-minimal coupling of the Higgs doublet $\zeta$ is much smaller than $\xi$.

Working in the Einstein frame, we first minimize the scalar potential with two fields.  For convenience in the following discussions, we enumerate the terms of the scalar potential (\ref{einpot2}) with $\zeta=0$ as follows,
\be
 V_E=V_0\left[1+a\Big(\frac{\phi}{\sigma}\Big)^4-2b\Big(\frac{\phi}{\sigma}\Big)^2 +2\frac{m^2_\sigma}{\sigma^2}+2m^2_\phi\frac{\phi^2}{\sigma^4}+\frac{c}{\sigma^4} \right]. \label{unitarypot2}
\ee
Compared to the scalar potential (\ref{einpot2}), we have chosen the parameters as $V_0=\frac{1}{4}\kappa \Lambda^2$,
$a=\alpha^2+\frac{\lambda}{\kappa}, b=\alpha$,  $m^2_\phi=\alpha\Lambda^2-\frac{\lambda}{\kappa}v^2$,
$m^2_\sigma=-\Lambda^2$ and $c=\Lambda^4+\frac{\lambda}{\kappa}v^4$. Then, at the minimum of the total potential, we can determine the Planck scale with a nonzero $\sigma$ vev and the electroweak scale with a nonzero Higgs vev.

From eq.~(\ref{unitarypot2}), 
for $\frac{\partial V_E}{\partial\phi}=\frac{\partial V_E}{\partial\sigma}=0$,
we obtain the minimization conditions
\bea
\sigma^2&=&\frac{a}{b}\phi^2+\frac{m^2_\phi}{b}, \label{minimum1}\\
\phi^2&=&-\frac{1}{m^2_\phi}(m^2_\sigma \sigma^2+c). \label{minimum2}
\eea
For $m^2_\phi=m^2_\sigma=c=0$, we find that there is a flat direction along $\sigma^2=\frac{a}{b}\phi^2$.

For the Einstein-frame action (\ref{einlag55}), from the formula (\ref{genepsilon}), we get the $\varepsilon$ parameter for the two-fleld inflation as
\be
\varepsilon\simeq \frac{M^2_P}{2V^2_E}\Big(\frac{\sigma}{\Lambda}\Big)^2
\bigg[\frac{\sigma^2-M^2}{\sigma^2+6\xi( \sigma^2-M^2)}\Big(\frac{\partial V_E}{\partial\sigma}\Big)^2
+\Big(\frac{\partial V_E}{\partial\phi}\Big)^2\bigg]. \label{epsilon0}
\ee
The contribution coming from the $\sigma$ derivative is suppressed by a large non-minimal coupling.
So it is reasonable to take the inflaton direction to be along the line 
with $\frac{\partial V_E}{\partial \phi}=0$, which is equal to $\sigma^2=\frac{a}{b}\phi^2+\frac{m^2_\phi}{b}$.
This inflaton direction corresponds to the flat direction in the limit of vanishing dimensionful parameters. 
For the inflaton direction, we simplify the potential and its derivatives, 
\bea  
V_E&=&\frac{V_0}{a}\Big[a-b^2+\frac{2}{\sigma^2}(bm^2_\phi+am^2_\sigma)+\frac{1}{\sigma^4}(ac-m^4_\phi)\Big],  \label{pot}\\
\frac{\partial V_E}{\partial\sigma}&=&\frac{4V_0}{a\sigma}\Big[-\frac{1}{\sigma^2}(bm^2_\phi+am^2_\sigma)+\frac{1}{\sigma^4}(m^4_\phi-ac)\Big], \label{dderiv1}\\
\frac{\partial^2 V_E}{\partial\phi^2}&=&\frac{8V_0}{a\phi^2}\Big(b-\frac{m^2_\phi}{\sigma^2}\Big)^2,
\label{dderiv2}\\
\frac{\partial^2 V_E}{\partial\sigma^2}&=&\frac{8V_0}{a\sigma^2}\Big[b^2+\frac{3}{2\sigma^2}(bm^2_\phi+am^2_\sigma)+\frac{5}{2\sigma^4}(ac-m^4_\phi)\Big],  \label{dderiv3}\\
\frac{\partial^2 V_E}{\partial\sigma\partial\phi}&=&\frac{8V_0}{a\phi\sigma}\Big(-b^2+b\frac{m^2_\phi}{\sigma^2}\Big).  \label{dderiv4}
\eea 

Then,  from eq.~(\ref{epsilon0}), using eqs.~(\ref{pot}) and (\ref{dderiv1}), the $\varepsilon$ parameter becomes
\be
\varepsilon\simeq \frac{4V^2_0}{3a^2V^2_E\sigma^4}\Big[bm^2_\phi+am^2_\sigma+\frac{1}{\sigma^2}(ac-m^4_\phi)\Big]^2.\label{epsilon1}
\ee

In order to compute the $\eta$ parameter for the scalar kinetic terms given in eq.~(\ref{einlag55}), we first consider the non-zero components of the Christoffel symbol 
\be
\Gamma^\sigma_{\phi\phi}=\frac{\sigma^2- M^2}{\sigma^2+6\xi( \sigma^2-M^2)}\frac{1}{\sigma}, \quad \Gamma^\phi_{\phi\sigma}=-\frac{1}{\sigma}=\Gamma^\sigma_{\sigma\sigma}.
\ee
Then, the matrix elements of $N_I\,^J$ in eq.~(\ref{generaleta}) are
\bea
N_\phi\,^\phi&=& \frac{M^2_P}{V_E} \Big(\frac{\sigma}{\Lambda}\Big)^2 
\Big(\partial^2_\phi V_E-\frac{\sigma^2-M^2}{\sigma^2+6\xi( \sigma^2- M^2)}\frac{1}{\sigma}\partial_\sigma V_E\Big), \label{N1} \\
N_\sigma\,^\sigma &=& \frac{M^2_P}{V_E} \Big(\frac{\sigma}{\Lambda}\Big)^2\frac{\sigma^2-M^2}{\sigma^2+6\xi( \sigma^2-M^2)}
\Big(\partial^2_\sigma V_E+\frac{1}{\sigma}\partial_\sigma V_E\Big), \label{N2} \\
N_\phi\,^\sigma&=&  \frac{M^2_P}{V_E} \Big(\frac{\sigma}{\Lambda}\Big)^2\frac{\sigma^2- M^2}{\sigma^2+6\xi( \sigma^2-M^2)}
\Big(\partial_\phi\partial_\sigma V_E+\frac{1}{\sigma}\partial_\phi V_E\Big),  \label{N3}\\
N_\sigma\,^\phi&=& \frac{M^2_P}{V_E}\Big(\frac{\sigma}{\Lambda}\Big)^2\Big(\partial_\sigma\partial_\phi V_E+\frac{1}{\sigma}\partial_\phi V_E\Big). \label{N4}
\eea
For the inflaton direction with large non-minimal coupling satisfying $6\xi\gg \frac{\sigma^2}{\sigma^2-M^2}$,  from eqs.~(\ref{N1})-(\ref{N4}) we obtain
\bea
N_\phi\,^\phi &\simeq & \frac{\xi\sigma^2}{V_E}\partial^2_\phi V_E   \\
N_\sigma\,^\sigma &\simeq & \frac{\sigma^2}{6V_E}\Big(\partial^2_\sigma V_E +\frac{1}{\sigma} \partial_\sigma V_E \Big)\\
N_\phi\,^\sigma &\simeq &  \frac{\sigma^2}{6V_E}\partial_\sigma\partial_\phi V_E, \nonumber \\
N_\sigma\,^\phi &\simeq & \frac{\xi\sigma^2}{V_E} \partial_\sigma\partial_\phi V_E.
\eea
Therefore, plugging eqs.~(\ref{pot})-(\ref{dderiv4}) in the above, we obtain the eigenvalues $\eta_1,\eta_2$ of the matrix $N_I\,^J$ as
\bea
\eta_1&\simeq & \frac{4V_0}{3a\sigma^2V_E}(bm^2_\phi+am^2_\sigma)\Big(1-\frac{m^2_\phi}{b\sigma^2}\Big)^2,  \label{eta01}\\
\eta_2&\simeq & \frac{8\xi V_0}{aV_E}\Big(b-\frac{m^2_\phi}{\sigma^2}\Big)^2\Big(\frac{\sigma}{\phi}\Big)^2. \label{eta02}
\eea

\end{document}